\documentclass{article}
\usepackage{spconf}
\usepackage{amsmath}
\usepackage{multirow}
\usepackage{afterpage}
\usepackage{siunitx}
\usepackage{rotating}
\usepackage{booktabs}
\usepackage{caption}

\usepackage{enumitem}

\usepackage[dvipsnames]{xcolor}

\setlist{nosep, leftmargin=14pt}

\usepackage{mwe} 

\title{3D Volumetric Super-Resolution in Radiology Using 3D RRDB-GAN}

\name{Juhyung Ha$^1$ \qquad Nian Wang$^{2,3}$ \qquad Surendra Maharjan$^{2}$ \qquad Xuhong Zhang$^1$}

\address{
    $^1$ Luddy Computer Science Department, Indiana University, Bloomington, IN USA \\
    $^2$ Department of Radiology and Imaging Sciences, Indiana University, Indianapolis, IN USA\\
    $^3$ Stark Neurosciences Research Institute, Indiana University, Indianapolis, IN USA 
}

\begin{document}
%
\maketitle

\section{Abstract}
\vspace{-0.1cm}
This study introduces the 3D Residual-in-Residual Dense Block GAN (3D RRDB-GAN) for 3D super-resolution for radiology imagery. A key aspect of 3D RRDB-GAN is the integration of a 2.5D perceptual loss function, which contributes to improved volumetric image quality and realism. The effectiveness of our model was evaluated through 4x super-resolution experiments across diverse datasets, including Mice Brain MRH, OASIS, HCP1200, and MSD-Task-6. These evaluations, encompassing both quantitative metrics like LPIPS and FID and qualitative assessments through sample visualizations, demonstrate the model’s effectiveness in detailed image analysis. The 3D RRDB-GAN offers a significant contribution to medical imaging, particularly by enriching the depth, clarity, and volumetric detail of medical images. Its application shows promise in enhancing the interpretation and analysis of complex medical imagery from a comprehensive 3D perspective.

\begin{keywords}
    3D super resolution, T1/ T2 MRI, CT, MRH, GAN, perception loss
\end{keywords}


\section{Introduction}
\vspace{-0.1cm}
Super-resolution (SR) is a crucial image processing task aimed at enhancing the resolution of images, making it an essential step for numerous applications. In the realm of medical imaging, the significance of SR is magnified as it directly impacts the diagnostic capabilities, thereby affecting clinical decisions. Achieving high-resolution (HR) imaging is especially critical when the target is to delineate fine anatomical structures and pathologies in three dimensions, facilitating a comprehensive understanding of the medical conditions under study.

The traditional methods of SR predominantly operate in two dimensions, which may not capture the full essence of volumetric data present in medical imaging. This necessitates the development of 3D SR techniques that can efficiently handle the inherent 3D nature of medical images, preserving and enhancing the information across the axial, sagittal, and coronal views.

In this work, we propose a 3D Residual-in-Residual Dense Block GAN (3D RRDB-GAN) aiming to generate realistic high-resolution volumetric medical images in the scaling factor of 4. Furthermore, we compared our methods to other 3D-based SR methods to test its capabilities over others existing methods. Our approach diverges from standard SR methods by introducing a 2.5D perceptual loss function designed to enhance the visual realism in all dimensions.


\afterpage{
    \begin{figure}
    \centering
    \includegraphics[width=1\linewidth]{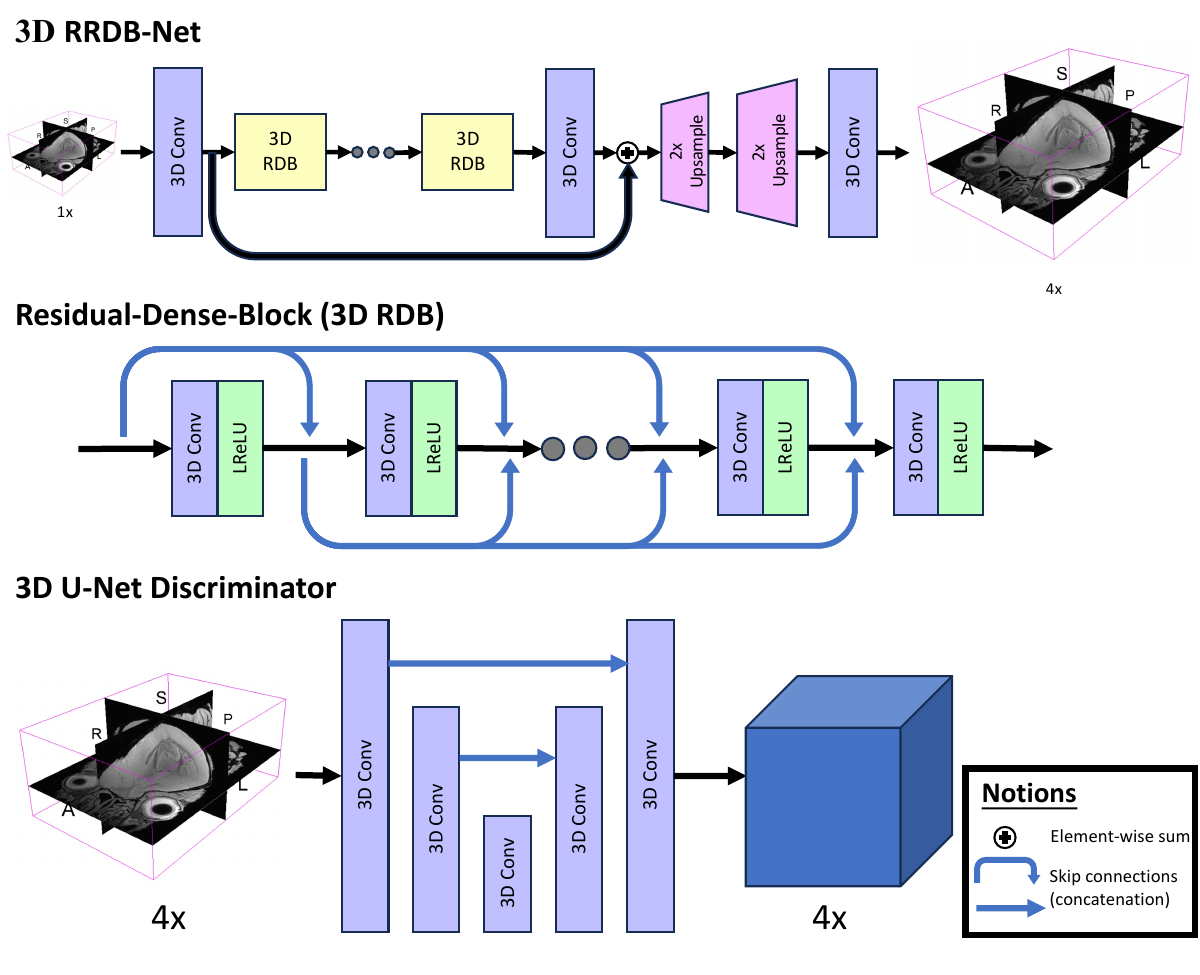}
    \vspace*{-0.5cm}
    \caption{Model architecture for 3D RRDB-GAN. Generator (3D RRDB-Net) uses stacks of 3D Residual-Dense-Block (3D RDB). 3D RDB has stacks of convolutional layers and dense-residual skip connections by concatenation. Discriminator is 3D U-Net discriminator outputs voxel-level classification with the same size of input.}
    \label{fig:model_arch}
    \vspace*{-0.4cm}
    \end{figure}
}

\section{Related Methods}
\vspace{-0.1cm}
\subsection{Deep Learning Approaches for Super Resolution}
Deep learning has transformed the field of super-resolution, offering models that can learn complex, hierarchical representations from the data. Convolutional Neural Networks (CNNs), for instance, have demonstrated superior performance in SR tasks over traditional methods \cite{7780550}. RRDBNet \cite{wang2018esrgan}, RCAN \cite{Lim_2017_CVPR_Workshops}, SRResNet \cite{8099502}, and DCSRN \cite{8363679} are landmark models that harness the power of CNNs for image upscaling. The introduction of Generative Adversarial Networks (GANs) further boosted the SR capabilities by enabling the generation of high-resolution images that are perceptually closer to real images \cite{8099502}.

\vspace{-0.1cm}
\subsection{2D vs. 3D Models in Medical Imagery}
While 2D Super-Resolution (SR) models have demonstrated efficacy in domains such as natural image processing, and have even found utility in enhancing medical images \cite{Yang2023-wz}, their application in 3D medical imaging scenarios like CT and MRI remains suboptimal. Predominantly, 2D SR models have been employed for image enhancement in the axial view, overlooking the depth dimension intrinsic to 3D medical imagery and limiting the resolution enhancement only in one 2-dimensional view. This oversight restricts their applicability in 3D radiology, given the volumetric nature of the data at hand. The advent of 3D Convolutional Neural Networks (3D CNNs) has marked a pivotal shift by incorporating the depth dimension, thereby providing a richer contextual framework and preserving spatial relationships within medical images \cite{10.1007/978-3-319-46723-8_49}. Unlike their 2D counterparts, 3D models endeavor to preserve and augment information across not only the axial but also the sagittal and coronal views, presenting a more holistic approach to medical image super-resolution.


\vspace{-0.1cm}
\subsection{Perception Loss}
Perceptual loss functions \cite{10.1007/978-3-319-46475-6_43, Wu2020} have been introduced to enhance the visual realism of synthetically generated images, including super-resolved images. In addition to pixel-wise accuracy, this loss function aims to generate perceptually meaningful results by focusing on maintaining structural and textural information in the SR images. Perceptual loss is typically deriveed from pre-trained networks such as VGG models. Due to lack of pre-trained models in 3D medical imaging, we utilized 2.5D perception loss by aggregation of 2D perception loss across 3 different directions (axial, sagittal, and coronal) using pre-trained 2D VGG model.

\section{Methods}
\vspace{-0.1cm}
\subsection{Model Development}
For the development of our super-resolution model, we utilized a 3D RRDB-GAN (Figure \ref{fig:model_arch}). The RRDB network has shown the state-of-the-art performance in 2D super resolution in natural images \cite{wang2021realesrgan}. By extending this network to operate in 3D, we are able to fully leverage the volumetric nature of medical imaging data. In addition, we introduce 3D UNet discriminator and 2.5D perceptual loss to ensure that the super-resolved images are perceptually similar to the high-resolution target images in 3D space.



\vspace{-0.1cm}
\subsection{Loss functions}
As we utilize GAN framework, our losses (eq.\ref{eq:total_loss}) includes generator's loss ( $L_{gen}$) and discriminator's loss ($L_{disc}$). 
\begin{align}
L=&L_{gen}+L_{disc} \label{eq:total_loss}\\
L_{gen}&=\lambda_1 L_{pixel} + \lambda_2 L_{perc} + \lambda_3 L_{adv} \label{eq:gen_loss}\\
L_{perc}&=\sum_{view}^{a,c,s}|VGG(I_{sr})-VGG(I_{hr})| \label{eq:perception_loss}
\end{align}
$L_{gen}$ (eq.\ref{eq:gen_loss}) contains 3 loss functions such as pixel-loss ($L_{pixel}$), 2.5D perceptual loss ($L_{perc}$), and adversarial loss ($L_{adv}$) with tunable weights. For our experiments, we used $\lambda_1=1$, $\lambda_2=1$, $\lambda_3=0.01$). We used mean-absolute-error (MAE) for $L_{pixel}$. For $L_{disc}$ and $L_{adv}$, we followed traditional binary classification loss functions, but in voxel-wise level using 3D UNet discriminator \cite{wang2018esrgan} (depicted as discriminator in Figure \ref{fig:model_arch}). Lastly, $L_{perc}$ (eq.\ref{eq:perception_loss}) is aggregated 2D VGG19 perceptual loss across every axial, coronal, and sagittal view, thus 2.5D. This 2.5D aggregation loss allows volumetric image quality enhancement.

\vspace{-0.1cm}
\subsection{k-space Image Degradation}
In the domain of super-resolution, the practice of image degradation to create paired low-resolution and high-resolution images is required. We adopted k-space image degradation method, aligning with the intrinsic nature of medical image acquisition via k-space. The low-resolution image through k-space degradation is attained through a series of steps: (1) applying 3D Fast Fourier Transform (FFT) to the 3D image, (2) center cropping the resultant 3D frequency matrix by the upscaling factor, (3) zero-padding the cropped 3D frequency matrix to match the original image dimensions, (4) reverting the frequency matrix into image space using 3D Inverse Fast Fourier Transform (IFFT), and (5) executing nearest-neighbor interpolation to downsample the image into a low-resolution version.

\vspace{-0.1cm}
\subsection{Evaluation setting}
The evaluation of the SR models was done by comparing the generated SR images against the original HR images. For a comprehensive evaluation, we employed four key metrics: Structural Similarity Index (SSIM \cite{Wang2004}), Peak Signal-to-Noise Ratio (PSNR), Learned Perceptual Image Patch Similarity (LPIPS \cite{Zhang2018}), and Frechet Inception Distance (FID \cite{heusel2017gans}). Traditional evaluation metrics such as SSIM and PSNR are used to evaluate structural similarities and fidelity between synthetic and real images. On the other hand, modern evaluation approaches such as LPIPS and FID captures the perceptual similarities between synthetic and real images by comparing high-level features. The main difference between traditional metrics (SSIM/PSNR) and modern approaches (LPIPS/FID) is that traditional metrics focus on pixel-level accuracy, where modern approaches focus on high-level feature representation, producing more human-like assessment of image quality \cite{Zhang2018}. Furthermore, we adopted 5-fold cross-validation to ensure the robustness and generalizability of our evaluation.


\afterpage{
    \begin{sidewaysfigure*}[h!tb]
    \hspace*{-1.5cm}
    \includegraphics[width=0.55\linewidth,height=0.35\textwidth]{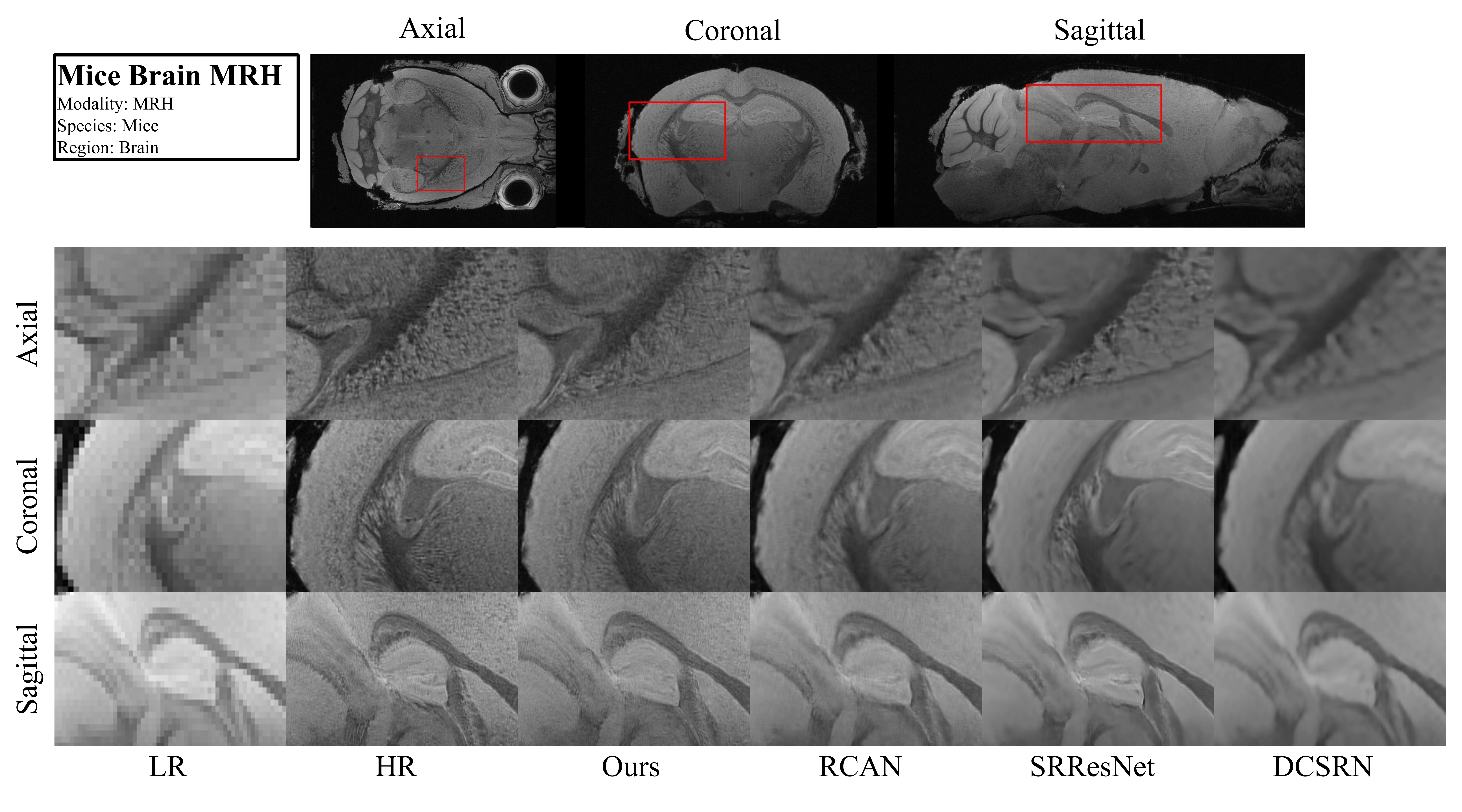}
    \includegraphics[width=0.55\linewidth,height=0.35\textwidth]{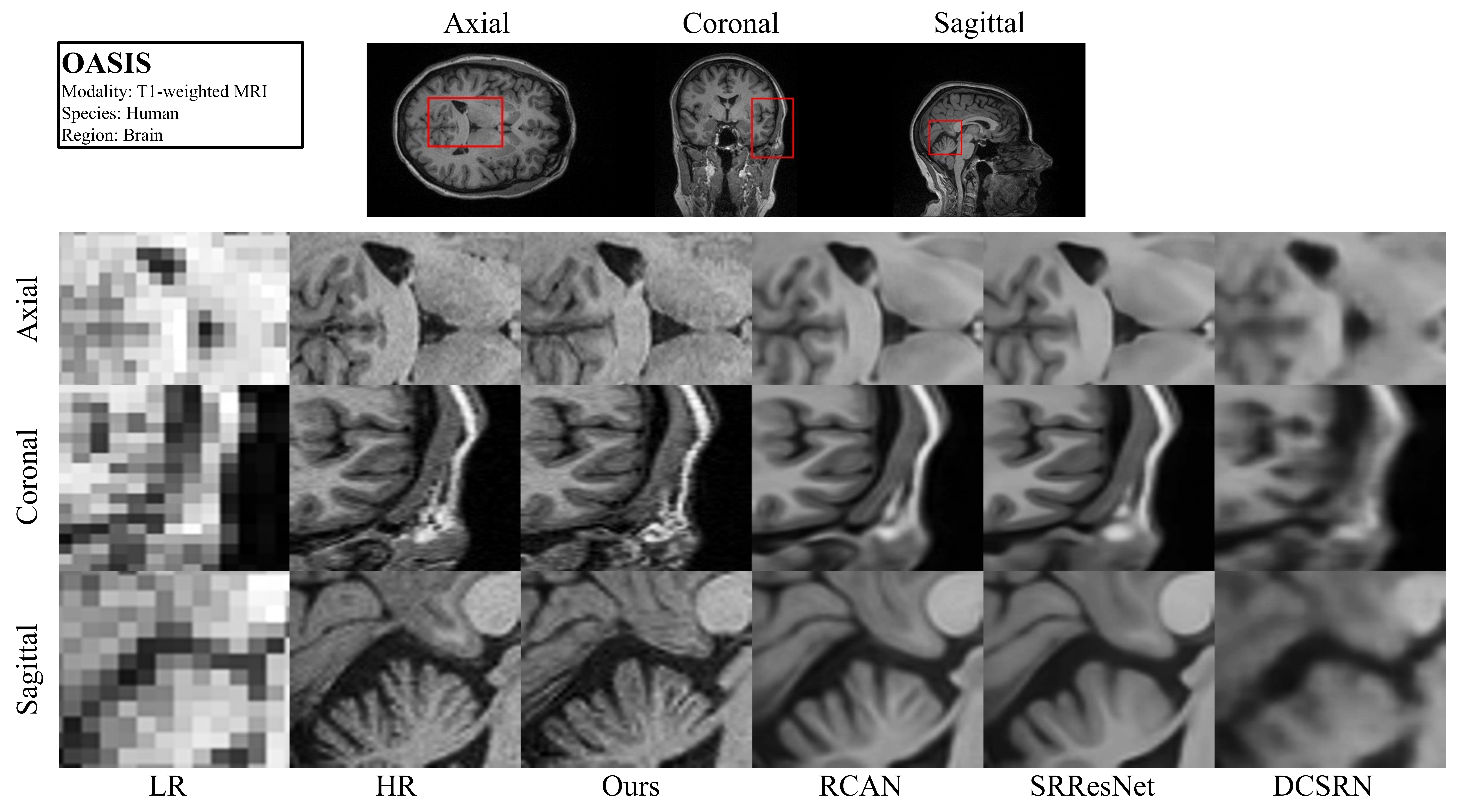}
    \hspace*{-1.5cm}
    \includegraphics[width=0.55\linewidth,height=0.35\textwidth]{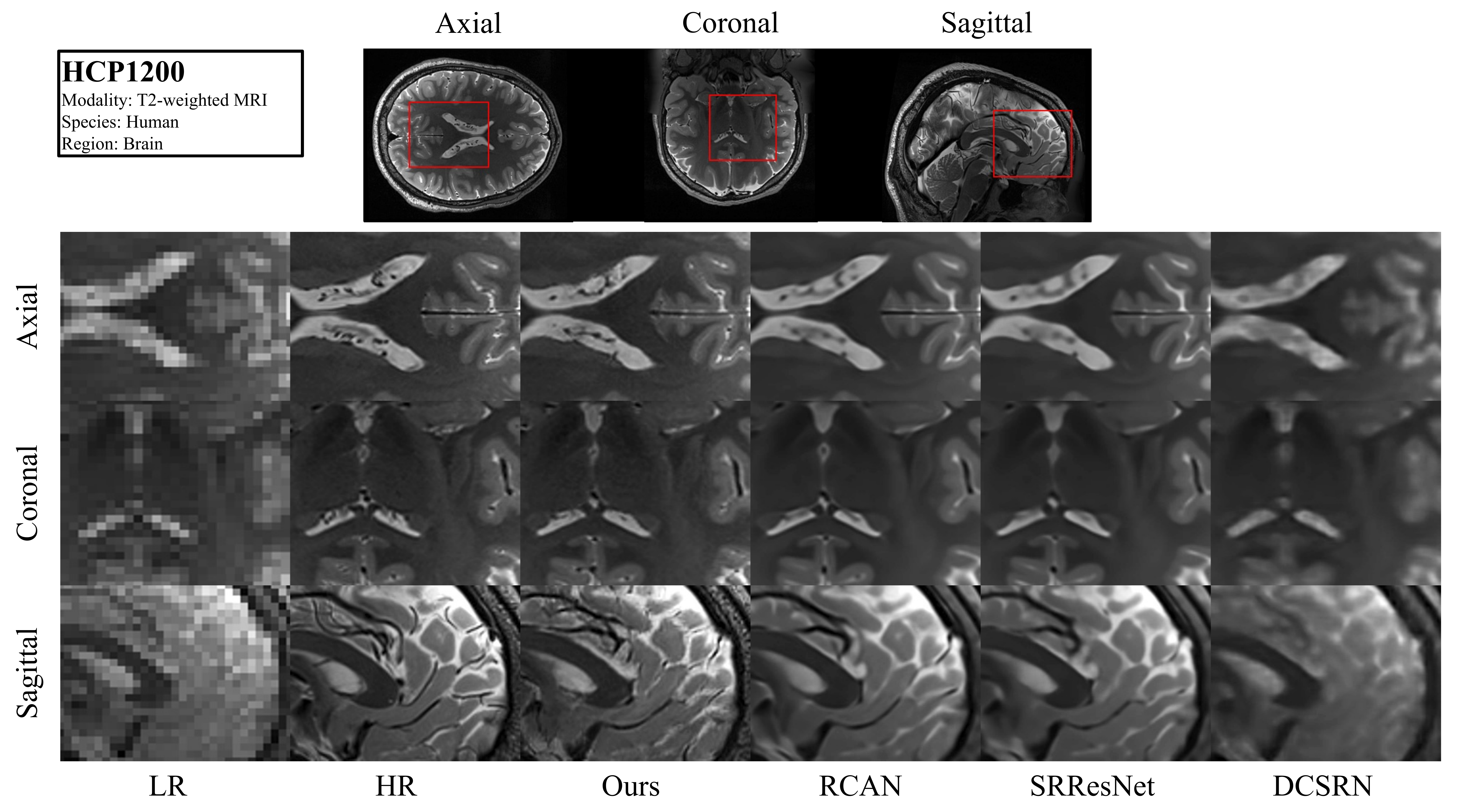}
    \includegraphics[width=0.55\linewidth,height=0.35\textwidth]{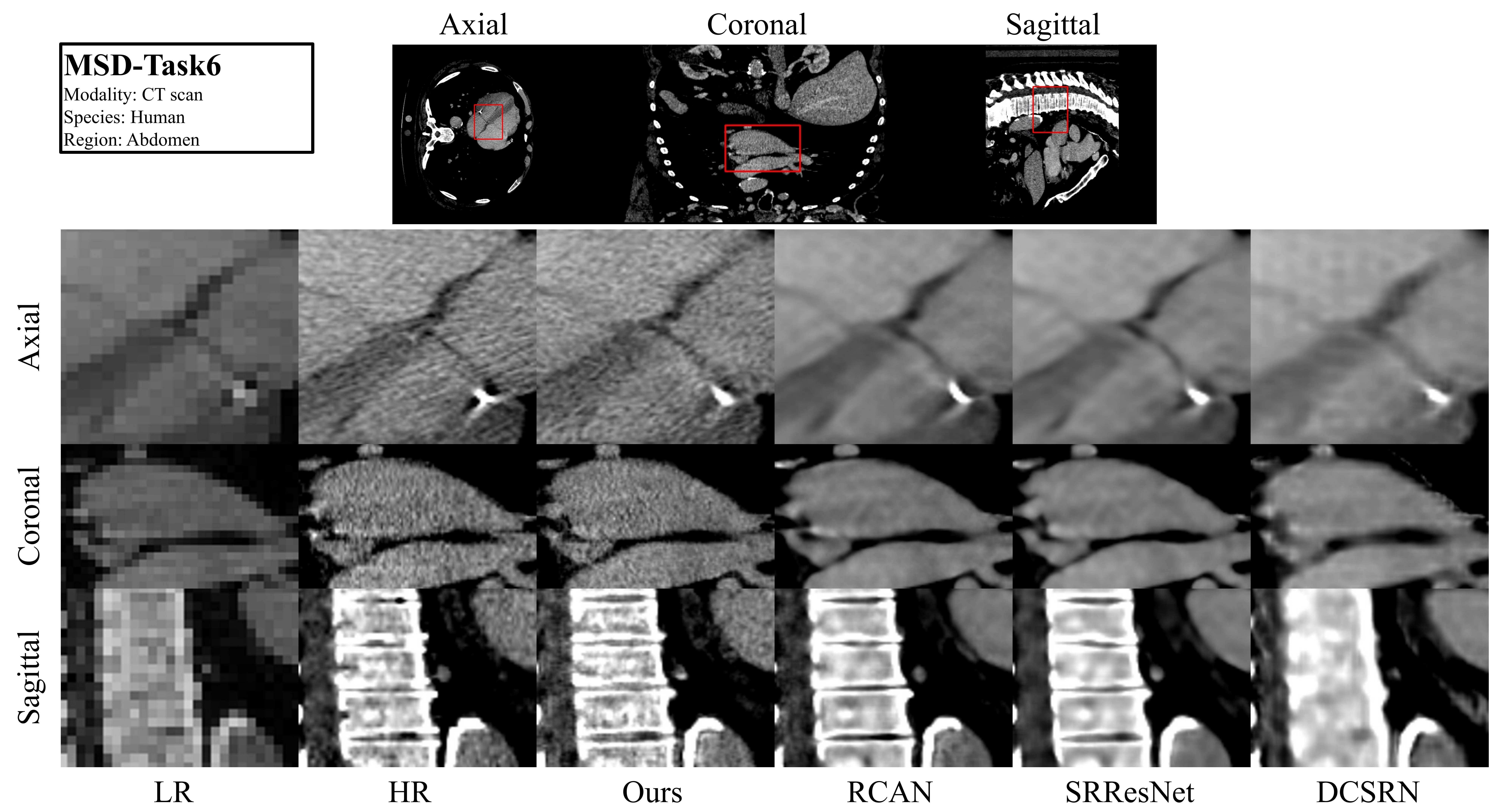}
    \captionsetup{margin=-1cm}
    \caption{Visualization results of a subject in 3D views (axial, coronal, and sagittal) from each dataset and method. Top row with 3 images show holistic views of subject, and 3 rows below show detailed view of zoomed-in patches in the red box in HR. Low-resolution (LR) was used as an input to the models. For visual purpose, trilinear interpolation was performed to LR to match the resolution with others. High-resolution (HR) is a ground-truth imaging, and others are generated by multiple deep-learning based models. For the CT scans (MSD-Task6), voxel values were clipped bewteen 0 and 250 to add more contrasts for organs and bones.}
    \label{fig:sample_figure}
    \end{sidewaysfigure*}
}
\vspace{-0.1cm}

\begin{table}[t!]
\centering
\scriptsize
    \vspace{-0.5cm}
  \begin{tabular}{p{1.4cm}SSSSS}
    \toprule
    {Dataset} & {Method} & {SSIM $\uparrow$} & {PSNR $\uparrow$} & {LPIPS $\downarrow$} & {FID $\downarrow$} \\
    \midrule
    \multirow{5}{*}{Mice Brain MRH}& {Trilinear} & {0.79} & {29.38} & {0.32} & {6.69} \\
    & {3D SRResNet \cite{8099502}} & \textbf{0.85} & \textbf{33.30} & {0.18} & {4.07} \\
    & {3D DCSRN \cite{8363679}} & {0.82} & {32.34} & {0.22} & {4.29} \\
    & {3D RCAN \cite{Lim_2017_CVPR_Workshops}} & {0.84} & {32.73} & {0.12} & {1.91} \\
    & {Ours} & {0.82} & {32.21} & \textbf{0.06} & \textbf{0.35}\\
    
      \midrule
      \multirow{5}{*}{OASIS}& {Trilinear} & {0.79} & {26.55} & {0.25} & {6.89} \\
    & {3D SRResNet} & {0.88} & {31.95} & {0.12} & {4.17} \\
    & {3D DCSRN} & {0.82} & {29.96} & {0.17} & {5.41} \\
    & {3D RCAN } & \textbf{0.89} & \textbf{32.27} & {0.11} & {3.69} \\
    & {Ours} & {0.82} & {29.50} & \textbf{0.04} & \textbf{0.18}\\
    
      \midrule
      \multirow{5}{*}{HCP1200}& {Trilinear} & {0.77} & {26.94} & {0.29} & {4.67} \\
    & {3D SRResNet } & {0.92} & {33.74} & {0.10} & {1.85} \\
    & {3D DCSRN} & {0.86} & {31.06} & {0.17} & {3.07} \\
    & {3D RCAN } & \textbf{0.93} & \textbf{34.09} & {0.10} & {1.60} \\
    & {Ours} & {0.88} & {31.61} & \textbf{0.04} & \textbf{0.09}\\

      \midrule
      \multirow{5}{*}{MSD Task 6}& {Trilinear} & {0.80} & {28.82} & {0.27} & {3.94} \\
    & {3D SRResNet } & \textbf{0.93} & \textbf{37.89} & {0.08} & {1.70} \\
    & {3D DCSRN} & {0.88} & {33.88} & {0.14} & {2.72} \\
    & {3D RCAN } & {0.92} & {37.79} & {0.07} & {1.38} \\
    & {Ours} & {0.89} & {36.01} & \textbf{0.03} & \textbf{0.18}\\
    \bottomrule
  \end{tabular}
  \vspace{-1mm}
\caption{3D super-resolution (4x upscaling) results from 4 datasets with 4 evaluation metrics. Trilinear is a baseline method with simple interpolation. 3D SRRestNet, 3D DCSRN, and 3D RCAN are comparison methods, and our proposed method (ours) is based on 3D RRDB-GAN. Best performed scores are marked with \textbf{bold} font.}
\vspace{-0.5cm}
\label{tab:result_table}
\end{table}
\vspace{-0.1cm}

\section{Results}
\vspace{-0.1cm}
\subsection{Datasets and Implementation}

In our study, four diverse datasets were employed: \textit{Mice Brain MRH} \cite{Wang2018,Wang2019, CRATER2022119199}, \textit{OASIS} \cite{10.1162/jocn.2007.19.9.1498}, \textit{HCP1200} \cite{HCP}, and \textit{MSD-Task-6} \cite{DBLP:journals/corr/abs-1902-09063}. The \textit{Mice Brain MRH} dataset comprises 11 high-resolution MRH images of mice brains, each with a voxel size of $25 \mu m^3$ and resolution of $512\times 720\times 304$. The \textit{OASIS} dataset includes 1,668 human T1-weighted brain MRIs, each with a voxel size of $1mm^3$ and a median resolution of $176\times256\times256$. From the \textit{HCP1200} dataset, which contains 1,068 human brain scans in multiple modalities, we utilized T2-weighted MRIs with a voxel size of $0.7mm^3$ and resolution of $260\times 312\times 260$. The \textit{MSD-Task-6} dataset features 60 human abdominal CT scans, with voxel sizes ranging from $0.6$ to $1mm$ for the axial and coronal axes, and $0.6$ to $2.5mm$ for the sagittal axis. To standardize the voxel size, images were spatially rescaled to a median voxel unit of $0.785\times 0.785\times 1.25$, resulting in a median resolution of $512\times512\times242$.

For our experiments, patch-wise training and inference were conducted using patches of size $96\times 96\times 96$, optimizing computing memory resources. This approach facilitates our model to perform super-resolution tasks scaling from $24\times24\times24$ to $96\times96\times96$. Weighted sampling was employed during training to consistently include regions of interest in the samples. Inference was executed using a sliding window approach with the Monai Python package \cite{cardoso2022monai}, maintaining the same patch size used in training. Each experiment utilized a single GPU from cluster nodes (NVIDIA A100 GPUs and AMD EPYC 7742 CPUs).


\vspace{-0.1cm}
\subsection{Experiment results}
In this study, we conducted 4x upscaling task to assess the capabilities of our super resolution model, the 3D RRDB-GAN. Our evaluation utilized four distinct datasets, encompassing two species (mice and humans), two anatomical regions (brain and abdomen), and four imaging modalities (MRH, T1/T2 MRI, and CT). The outcomes of these evaluations are systematically presented in Table 1, where we compare various super-resolution methods.

When considering traditional metrics such as SSIM and PSNR, both 3D SRResNet and 3D RCAN demonstrated strong performance. However, our 3D RRDB-GAN model exhibited notable superiority in modern evaluation metrics, including LPIPS and FID, across all experimental datasets. Beyond these quantitative results, we also provide visualizations of sample cases from each dataset in Figure \ref{fig:sample_figure}. These visualizations offer a more intuitive comparison from a human perception standpoint, particularly evident in the zoomed-in regions where detailed imaging characteristics are highlighted.

\section{Conclusion}
\vspace{-0.1cm}
Our study demonstrates the efficacy of the 3D Residual-in-Residual Dense Block GAN (3D RRDB-GAN) in enhancing the quality of 3D medical images through super-resolution techniques. Our approach, distinguished by the integration of a 2.5D perceptual loss function, has shown its capability in significantly improving the volumetric detail and realism of images across various datasets. While our approach showed more pronounced results in visualizations and modern evaluation metrics like LPIPS and FID, indicating a strong perceptual quality, it encountered challenges in traditional metrics such as SSIM and PSNR. These outcomes highlight the specialized capability of the 3D RRDB-GAN in enhancing perceptual aspects of medical imaging, suggesting its potential utility in scenarios where detailed visual interpretation is crucial.

The breadth of datasets used in our study, covering different species, body regions, and modalities, underscores the versatility and adaptability of the 3D RRDB-GAN. While our model represents a significant step forward in medical imaging technology, future work will aim to explore its potential in broader clinical applications and its integration into diagnostic workflows. By continuously refining and adapting our model, we aspire to contribute further to the advancements in medical imaging, striving towards more accurate and efficient analysis and interpretation of complex medical data.

\newpage
\clearpage
\section{Acknowledgments}
\vspace{-0.1cm}
\label{sec:acknowledgments}
This work was supported in part by Indiana University Internal Research Seed Fund, and in part by the NIH R01 NS125020, Indiana Center for Diabetes and Metabolic Diseases Pilot and Feasibility Grant, Roberts Drug Discovery Fund \& TREAT-AD Center Grant.
\bibliographystyle{IEEEbib}
\bibliography{strings,refs}

\end{document}